\documentclass[12pt,a4paper]{article}

\usepackage{psfig}

\begin{document}

\thispagestyle{empty}

\title{\bf Implicit Simulations using Messaging Protocols}

\author{G.A. Kohring \\
        C\&C Research Laboratories, NEC Europe Ltd.\\
        Rathausallee 10, 53757 St. Augustin, Germany\\
        E-mail: kohring@ccrl-nece.de}

\date{}

\maketitle

\bigskip

\begin{abstract}
A novel algorithm for performing parallel, distributed computer 
simulations on the Internet using IP control messages is introduced.
The algorithm employs carefully constructed ICMP packets which 
enable the required computations to be completed as part of the standard 
IP communication protocol.  After providing a detailed description of the
algorithm, experimental applications in the areas of stochastic neural 
networks and deterministic cellular automata are 
discussed.  As an example of the algorithms potential power, 
a simulation of a deterministic cellular automaton 
involving $10^5$ Internet connected devices was performed.
\end{abstract}

\bigskip


\bigskip

\section{Introduction}

Most readers will be familiar with the concept of Moore's law as applied to
the computer industry, namely, the density of transistors on a chip and with 
it the speed
of the processors doubles roughly every 18 months.\cite{moore65} 
It has held true for more than three decades now and shows all signs 
of holding at least through the end of this decade. However, most readers will
be less familiar with similar growth laws in other areas of computer technology,
e.g., disk capacity has been doubling roughly every 12 months and
network bandwidth has been doubling roughly every 9 months.\cite{foster02} 
A mere 15 years ago the Internet backbone ran
at 64 Kbits/s, this year it is being upgraded to 40 Gbits/s, an increase of
nearly six orders of magnitude.
The latter
point is of particular interest for computational science, because faster
networks enable wide area distributed computing on a scale which was not
feasible in the past. However, it also implies that the same questions of
efficiency now afflicting the processors in modern PCs will soon become 
apparent in the communications networks themselves.  

As has been often mentioned,\cite{korpela01}
most processors are idle or running below their maximum capability most 
of the time,
with enormous wastes of cpu cycles. Projects like SETI@HOME\cite{korpela01} and 
Condor\cite{basney99} attempt
to harness this excess computing power for performing useful work. SETI@HOME's
approach is to hide their calculations behind screen savers which are
activated whenever the processor is idle.  Condor, on the other hand attempts
to make all the PCs and workstations in an organization available for running
parallel applications, whereby the load on each PC from a Condor job is 
adjusted to suit the needs of anyone accessing the PC interactively.

In the
same vane, the rapidly increasing bandwidth implies that most of a network's
capacity will soon go underutilized most of the time.  Hence, the question
to be addressed here, is whether or not the idle bandwidth can be tapped for
performing useful calculations.

Recently, Barab\'asi et al. examined this problem by utilizing the 
Hyper Text Transmission Protocol (HTTP)
to perform calculations aimed at solving a 2-SAT problem during the 
act of communicating with web servers.\cite{barabasi01}
Although their algorithm worked as expected, it has several drawbacks. 
1) The HTTP\cite{rfc1945,rfc2616}
protocol uses {\it connected sockets},  which require three 
messages to be
exchanged in order to initiate the connection before any data can be sent. 
Once connected, their algorithm requires two further message exchange wherein
the actual computation is performed. Afterwards another message has to be
exchanged to properly disconnect the sockets. In total, six messages are 
exchanged for each computation step.
2) The HTTP protocol is layered on top of the Transmission Control Protocol
(TCP)\cite{rfc793} which is in turn
layered upon the Internet Protocol (IP).\cite{rfc791,rfc2474}
Since each layer introduces its own overhead, it
would be more efficient to use a protocol  which is closer to the IP level.
3) The number of HTTP servers is actually quite small compared to the total
number of devices connected to the Internet, therefore limiting the application 
to HTTP servers greatly reduces the pool of devices available for performing
useful work.

In this paper we describe a different approach based upon the 
Internet Control Message Protocol (ICMP)\cite{rfc792},
which is layered directly on top of IP. ICMP is a connectionless protocol,
meaning only the messages containing the data need to be exchanged. No 
control messages are required for establishing and ending the connection.
Furthermore, ICMP is required to be
implemented by every Internet capable device,  hence there are many more 
devices which can respond to ICMP messages than can respond to HTTP requests.
The next section describes the general approach in more detail. Section
three then discusses a version of the algorithm for stochastic neural 
network models
and section four describes an implementation for a deterministic cellular 
automaton.
We conclude with a discussion on the utility of these approaches and their
future viability.

\section{Computing with ICMP}

To understand how messages can be used to perform useful calculations one has
to understand how message exchange on the Internet works. 
The Internet is a so-called
packet-switching network,\cite{rfc792} meaning that all messages are broken down
into one or more packets called datagrams, whereby the order and meaning 
of the bytes within
the datagram are governed by the various transmission protocols which the
applications sending the messages are using.  Since the communication channels
used to transmit the packets are inherently noisy and unreliable, each
networking protocol adds some means, however rudimentary, of checking 
for corrupted data.  In the case of IP or ICMP, this takes the form of the
so-called {\it Internet check-sum} ($I_s$). If $I_s$ is inserted by the sender
into the original packet and then recalculated by the receiver, it is 
possible to
determine whether or not an error has occurred during transmission.  The basic
idea behind the current proposal is to exploit the calculation of $I_s$ to
perform useful work in addition to its primary function of checking 
whether or not the packet is corrupted.

Since the Internet Control Message Protocol is layered 
on top of IP, an IP/ICMP datagram consists of an IP header with 
20-60 bytes of control and routing information, plus the ICMP message itself. 
Most ICMP message types place severe restrictions on the type of data which 
can be sent as part of the message, however, the {\it Echo Request} and
{\it Echo Reply} messages
do not place any restrictions. Normally, this flexibility is used for
example by the {\it ping} command to measure bandwidth by measuring 
the time needed
for a datagram of a given size to complete a round-trip between two hosts,
or to check for size dependent transmission problems on a network connection.

\begin{figure}[htbp]
\centerline{\psfig{file=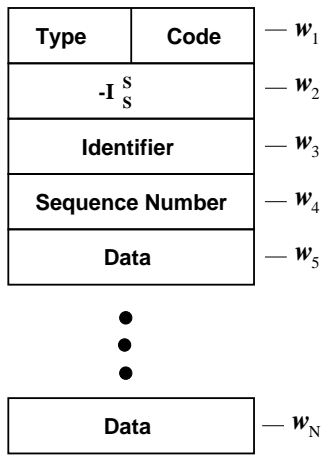,height=6.0cm}}
\caption{\label{fig:icmp}Format of an ICMP Message.}
\end{figure}

For our purposes, the bytes of an ICMP {\it Echo Request}
can be thought of as an array of $N$ 16-bit words as depicted in 
Fig.~\ref{fig:icmp}.
Whereby, the first byte of the first word indicates the ICMP message type
(8 for {\it Echo Request} and 0 for {\it Echo Reply}) and
the second byte of first word provides room for any special codes 
associated with the message type (always 0 for {\it Echo Request} and {\it Echo
Reply}). 
The second word contains the Internet
check-sum, $I_s^s$ for the ICMP part of 
the datagram, while the third and fourth
words contain information to help identify the datagram.
The remaining $N-4$ words comprise the data associated with the 
{\it Echo Request} and {\it Echo Reply}.

When computing the check-sum the 16-bit words are assumed to represent
integers in the 
{\it one's-complement} representation.\cite{rfc1071} In a one's-complement
representation negative integers are represented by inverting each bit
in the representation of their magnitude.  Curiously, this leads to
two representations for $0$, namely, $0...0$ and $1...1$, which are
designated $+0$ and $-0$ respectively.  (Contrast this with a two's complement
representation used on most computers where there is only one representation
for $0$ and negative numbers
are represented by subtracting one from the magnitude before inverting all
the bits.)

One's-complement addition, $\oplus$, is defined as
adding two numbers and carrying any overflow bit
around to the lowest order bit where it is added to the previous sum. This 
guarantees that
$1\oplus(-0)\equiv 1$ and not $+0$. Furthermore, the addition is circular,
i.e., if $M$ is the largest maximum integer, then $M\oplus1=-M$. With this
definition of addition, the Internet check-sum, $I_s$, is 
defined by the equation:

\begin{equation}
I_s = \bigoplus_{i=1}^{N}w_i,
\label{eq:check_sum}
\end{equation}
whereby, the sender sets $w_2=0$, for the purpose of calculating the 
initial checksum.
To determine whether or not the packet has been corrupted in transit, the
receiver computes the function:

\begin{equation}
F(I_s) = \delta_k(I_s,-0)
\label{eq:check_fun}
\end{equation}

\noindent where $\delta_k(x,y)=\{1\ {\rm if}\ x=y,\, 0\ {\rm otherwise}\}$ is 
the Kronecker delta function. If $F(I_s)=1$, the datagram is considered
to be uncorrupted. (To understand why $I_s=-0$ and not $I_s=+0$ indicates 
an uncorrupted packet, note that $1\oplus(-1)=-0$.)

For checksum purposes a one's complement arithmetic is preferable to a two's
compliment because its sensitivity to errors is independent of the 
bit position. In a two's compliment representation, flipping the
most significant bit in an even number of words would yield the same
checksum.\cite{rfc1071}

Obviously, this algorithm is not completely failsafe, since
the probability that any two random configurations will yield the
same value of $I_s$ is $2^{-16}$. However, it does guard against single bit
transmission errors and as these occur on average once in every
$2^{10}$ packets,\cite{stone98} it is sufficient for most applications. 
Applications requiring a lower error rate need to use a higher level
transmission protocol which more error checking.

When the receiver detects a corrupted packet 
it should simply discard it {\it without} emitting an error message to 
the sender. If the packet is not corrupt, then the receiver should respond
to an {\it Echo Request} with an {\it Echo Reply}.  When constructing the {\it Echo
Reply}
message, the receiver takes the original message, replaces the Type byte with
$0$ for {\it Echo Reply}, recomputes the checksum as if it were the sender and sends
the entire message back to the original sender, i.e., only $w_1$ and $w_2$ of
the original message are changed.
Essentially, then the {\it Echo Request} allows us to compute 
eq.~\ref{eq:check_fun} on the remote computer.

Normally, for higher level protocols like FTP, HTTP, etc., the job of 
calculating the checksum 
is done automatically by the software which implements the protocol so the
user never has to worry about it. However, for low level protocols like ICMP,
users are expected to construct the datagram, including $I_s$, themselves, 
which leads us to the
possibility of constructing $I_s$ in such a way that evaluating 
eq.~\ref{eq:check_fun} has a purpose other 
than that originally intended one of error checking.

In the definition of the ICMP there are a few small 
loopholes,\cite{rfc792}
e.g., the standards document does not explicitly state that the checksum
{\it must} be evaluated before answering an {\it Echo Request}.
In fact, some implementations, which we term {\it non-validating}, respond 
to an {\it Echo Request} without 
bothering to validate the checksum first.
In the same vane, the original sender is not explicitly required to validate
the {\it Echo Reply} response, thus we are at liberty to skip this time 
consuming step. (Strictly
speaking both of these actions are incorrect, for if the checksum is wrong,
the receiver
cannot be certain that the sender really meant to issue an {\it Echo Request}, 
or {\it Echo Reply} and not some other ICMP message type.
Nevertheless, these ``features'' will become very useful in what follows.)

\section{Simulating Stochastic Neural Networks}

Unlike HTTP, ICMP is not a reliable protocol, meaning there is no guarantee 
that a datagram will arrive, or that a reply will be sent,
or that replies will be sent in the same order as the requests, 
or that multiple replies will not be sent for each request or that the
requests will even be sent out in the first place. 
It is the job of a higher level protocol
to make these types of guarantees.  However, neural networks possess a built-in
robustness to noise, which renders these problems practically mute.

For the purposes of demonstrating the algorithm, we use a Hopfield model with
limited precision weights.\cite{hopfield82,kohring91,draghici02} (This is 
not a very serious restriction
because more complicated models can be incorporated using complex neurons with
a Hopfield type internal structure.\cite{kohring02})

Very briefly, the model consists of $N$ neurons denoted by $S_i\in\{-1,1\}$
and $N^2$ couplings denoted by $J_{ij}\in[-L,L]$, where $L$ is the maximum
allowed value for each coupling. Typically, $L$ only needs to be  a few
bits wide (4-5) for obtaining good results.\cite{draghici02,kohring91}
Finally, the model is endowed with a discrete, time dependent dynamics 
given by:

\begin{equation}
S_i(t+1)={\rm sign}\left(\sum_{j\not=i}^N J_{ij}S_j(t)\right).
\label{eq:neuro_model}
\end{equation}

\noindent The couplings, $J_{ij}$, are created using the clipped Hebb rule:

\begin{equation}
J_{ij}=B\left(-L,\, \sum_{\mu=1}^P \xi_i^{\mu}\xi_j^{\mu}, \, L\right)
\label{eq:hebb}
\end{equation}

\noindent where the $\{\xi^{\mu}\}$, with are the patterns the network should
learn and $B(a,x,b) = \{a\ {\rm if}\  x<a,\, b\ {\rm if}\  x>b, \, x \ 
{\rm otherwise}\}$

From eq. \ref{eq:neuro_model} it is evident that the computationally
intensive part of this simulation is the calculation of the sums.
By sending an ICMP message to a non-validating IP device, we can induce it to
compute this sum for us. 

First note that if we set the Identifier and Sequence Number to zero (refer
to fig.~\ref{fig:icmp}) and 
recall that in an {\it Echo Reply} message both Type and Code are zero, then we
see that the value of $I_s$ returned by the {\it Echo Reply} will be simply the sum
of the data elements.

For the {\it Echo Request} message we can set $I_s$ to any arbitrary value since
the non-validating IP implementation will just ignore it. The data elements
are then set to $J_{ij}$ or $-J_{ij}$ (in one's compliment representation) 
depending upon the value of $S_j(t)$.

Typically, a neural network is updated  either sequentially or in parallel,
although, Hopfield in his original work used an asynchronous
updating algorithm.\cite{hopfield82} In this paper we also use an asynchronous  
updating algorithm since it is a natural fit to the unreliability
of IP/ICMP messages mentioned above. After a message is sent 
a new message is constructed without waiting for a reply from the first
message. Once a reply has been received, the corresponding value
of $S_i$ is immediately updated.

We have tested this algorithm on our laboratory's intranet.  As a
first step, we selected a subset of devices which are non-validating.  None of
the workstations or PCs were in this subset, rather it consisted of printers,
routers and switches -- devices one would not normally think of using for
performing numerical simulations. The computer chosen for controlling the
simulations was an SGI O2 with a 180 MHz MIPS 5000 processor, while our
laboratory's intranet has a peak bandwidth of 100 Mbits/s.

\begin{figure}[htbp]
\centerline{\psfig{file=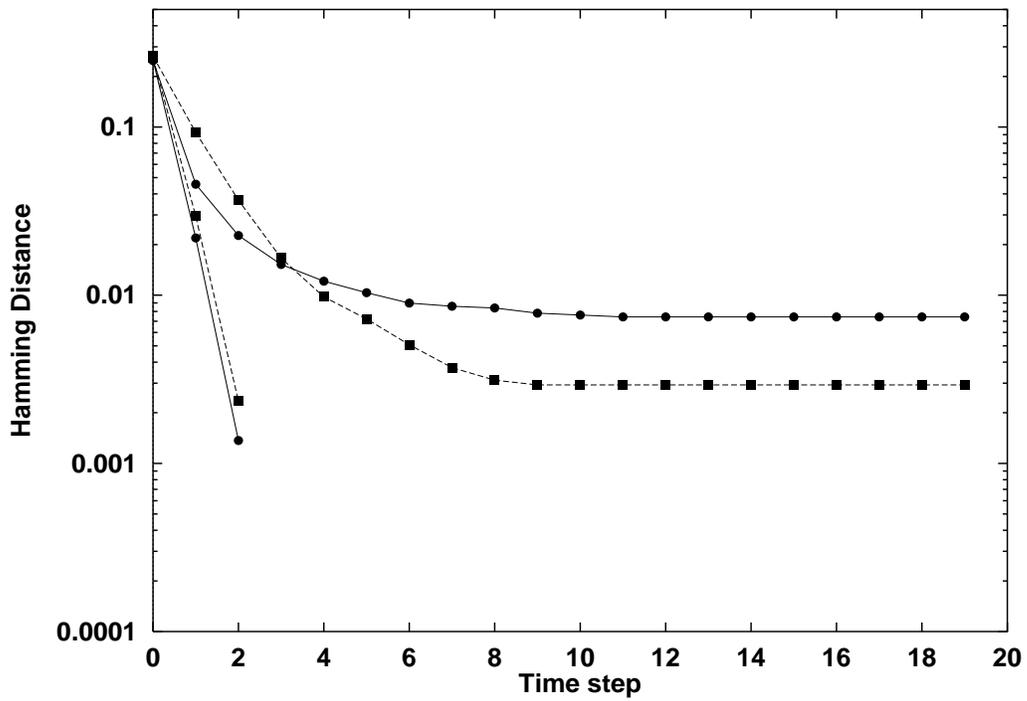,width=14.0cm}}
\caption{\label{fig:neuro}Hamming distance as a function of time for
a network of $N=512$ neurons. The solid lines represent standard parallel
updating while the dotted lines represent the present algorithm.
The upper curves are for $P = 48$
and the lower curve is for $P = 32$. The initial Hamming Distance is
0.25 and all data is averaged over ten
sets of patterns with ten patterns from each set. }
\end{figure}

A comparison of the present algorithm versus a standard parallel updating
algorithm for the same system size is shown in fig. \ref{fig:neuro}.
In these
experiments, we intentionally start with an initial configuration at a
Hamming distance of 0.25 from a learned pattern and observe how the standard
algorithm and the present network algorithm converge towards a stable state. 
(For the purposes of comparison, one time step in the asynchronous updating
algorithm corresponds to $N$ neural updates, while one time step in the
parallel updating consists of updating {\it all} $N$ neurons.)
Evidently, the quality of recall is not adversely affected by the asynchronous 
updating algorithm nor by the unreliability of the ICMP protocol.

We then tested the same algorithm on a set of randomly chosen devices residing
on the world-wide Internet.  Our results were similar to those in 
fig.~\ref{fig:neuro}, although the simulation ran much slower because 
our connection to the Internet has a bandwidth of only 1 Mbit/s. (We will delay
a more detailed discussion of the relative performance until section 5.)

These simple proof-of-concept experiments demonstrate that stochastic 
models can be simulated quite well using only
the computational capabilities of the messaging protocols themselves.
In the next section we discuss the algorithmic changes need 
to ensure that deterministic models can also be accurately simulated.

\section{Simulating Deterministic Cellular Automata}

For stochastic models, the unreliability of ICMP mentioned above can
be simply ignored, however, for deterministic systems, ICMP
has to be made reliable in much the same way higher level protocols
like TCP make IP reliable.

The model we chose for demonstration purposes is the
simple cellular automaton known as Life, which was invented by 
J.H. Conway.\cite{gardner70} This widely studied automaton is an 
interesting testing
ground because of its myriad properties, including the capability of
performing universal computations.\cite{berlekamp82}
In this model, the cells, $C_{ij}$, of a 
two dimensional square
lattice are initially assigned values $C_{ij}\in\{0,1\}$. The cells then change
their values in parallel by summing over the states of their nearest neighbors:

\begin{equation}
C_{ij}(t+1)=H\left(\sum_{m=i-1}^{i+1}\sum_{n=j-1}^{j+1}C_{mn}(t),\, 
C_{ij}(t)\right).
\label{eq:life_model_a}
\end{equation}

\noindent where,

\begin{equation}
H(x,y)=\cases{1& if \ $x-y=3$,\cr
              y& if \ $x-y=2$\cr
              0& otherwise.
             }
\label{eq:life_model_b}
\end{equation}
\noindent

For this experiment we want to make use of a large number of devices, hence, we
need an algorithm that works for validating devices as the number of
non-validating devices is much too small. As eq. \ref{eq:life_model_b} suggest,
the simplest approach is to send out two {\it Echo Request} for each cell, 
asking
if $x-y=2$ or $x-y=3$. To do this we copy the value $-w_1$ into the first
data element of the ICMP message (to counter the type value in the first
word),
then the  nearest neighbor cell values
are copied into the next 8 data elements of the ICMP message, followed by
a $-2$ ($-3$) to have the validating device determine if $x-y$ 
is 2 (3). If $H(x,y)$ does not evaluate to $1$ then the checksum will
be invalid and no response will come from the device.

To handle the unreliability  of ICMP, we adopt the following procedure.
First, out-of-order replies can be readily dealt with, if the time 
value, $t$, is placed in
the Identifier word, $w_2$. ($-t$ is then placed in the Sequence Number, $w_3$, 
word to balance the checksum.) Second, to guard against valid packets being
dropped because the datagram was corrupted during transit, the {\it Echo
Request} 
packets are first
sent out for all cells in the automaton, then the replies are examined. If a
given cell has not replied, a second {\it Echo Request} is sent. If again it does
not reply, then we conclude that the answer is indeed $H(x,y)=0$. (One, could
ask a third time, however, we find that two requests are generally
sufficient for accurate results.)

One might worry, that by using the checksum to evaluate eq.
\ref{eq:life_model_b} we are robbing the checksum of its primary function,
namely, detecting errors in transmitted packets.  Could for example, a single
bit error lead to a false positive? There are 3 factors to
be considered: 1) The probability of a single bit error occurring anywhere
in a datagram is less than $2^{-20}$.\cite{stone98} 2) Since we are only
asking if the checksum evaluates to 2 or 3, only errors in the 2 least
significant bits are important. 3) Of the $2^8$ possible
configurations of the neighboring cells, only half of them would yield a
different value of $H(x,y)$ if a single cell value were changed. Hence,
a false positive should occur less than 1 in $2^{24}$
cell updates - which is much more than we can achieve with this
algorithm using current networks.

As a first experiment involving the network automaton we created a 
small $4\times4$ automaton and
initialized it with a so-called {\it glider}\cite{gardner70} - a repeating
configuration of cells which move about in the space of the automaton. 
Then a control
automaton, implemented in the conventional fashion on a single processor was
given the exact same initial state and the two automata were left to
evolve, with a cell-by-cell comparison being made after each automata update.
During the simulation time of $2^{17}$ site updates, no deviation from the
control automaton was detected. In other words, the simple control procedures
explained above were enough to enable our deterministic automaton to run
using unreliable ICMP messages.

\begin{figure}[htbp]
\centerline{\psfig{file=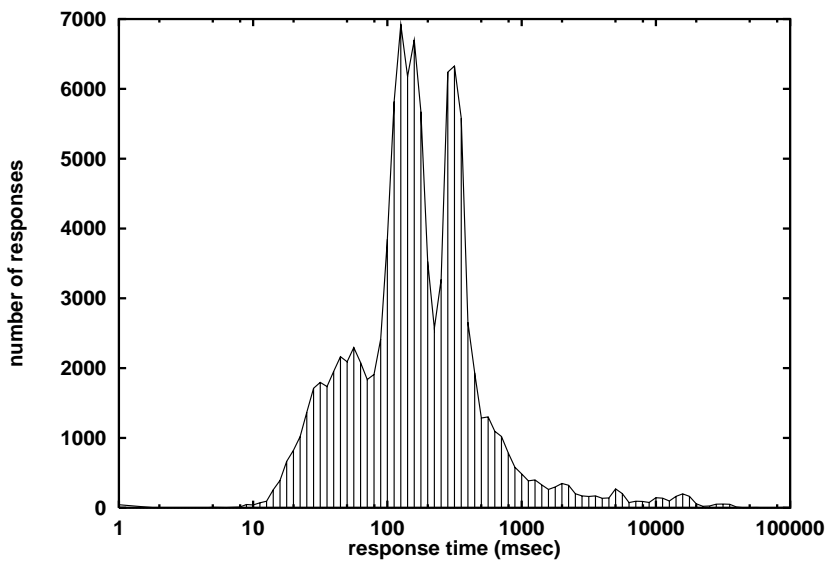,width=11.0cm}}
\caption{\label{fig:distrib}Histogram of average response time for an
{\it Echo Request} to a randomly chosen Internet device.}
\end{figure}

For our final experiment, we attempted to simulate an automaton with
$200\times 500$ cells using $10^5$ Internet devices. Although we expected this
experiment to be difficult given that our laboratories has only a 1 Mbit/s
connection to the Internet, an unexpected problem we
encountered here was the widely varying response times of the different
devices. Fig. \ref{fig:distrib} depicts the average response times 
for more than $10^5$ Internet devices. As can be seen, the response time
varies from a few milliseconds to several tens of seconds. In order to use 
as many of these devices as possible, we first discard all of those with an
average response time of more than 10 seconds. Then we order the remainder
from slowest to fastest and assign them sequentially to the cells beginning 
with $(1,1)$.  If we then send the requests sequentially to the
Internet devices, and delay examining the responses until all the requests
have been sent, we can mask the response time of the slowest devices. 
In this manner we were able to successfully update this very large system
for a few time steps. Unfortunately, the Internet is currently not 
stable enough for
simulations of this size and durations, therefore the experiments usually 
ended after two time steps because the Internet automata developed large 
deviations from the control automata.

\newpage

\section{Discussion}

The experiments described above have proven that the concept of computing
using communication is realizable, in other words,
it is possible to perform some types of simulations using
Internet messaging protocols. Beyond the proof-of-principle demonstrations 
are questions of performance and efficiency. As was stated in the
introduction, communication performance is increasing more rapidly than
computational performance. Currently, however, most networks are operating
far below state-of-the-art capabilities.

In the example of Conway's Life, we expected that the extra care needed for
making the communications reliable would reduce the efficiency of the
algorithm.  This was compounded by the firewalls set-up to protect our
laboratories computers from hackers that reduced our effective bandwidth to
the Internet from 100 Mbit/s to less than 1 Mbit/s. Given that a complete 
IP/ICMP datagram
including the IP header information for our automaton consists of 48 bytes,
we would expect to be able to send out a maximum of $2\,600$ messages per
second. In reality, we were barely able to sustain a speed of 200 messages
per second, meaning it took nearly $2\,000$ seconds to  complete one update
of our largest system. By contrast, it took our conventional program less than
half of one second to update the same system.  On our laboratories intranet, it
was possible to sustain a speed of approximately $1\,200$ cell updates per
second, which is still about 4 orders of magnitude slower than what can
be achieved on a single processor.

For the stochastic neural network models, the performance was far better.
Given a network of 512 neurons, each datagram, including the IP header,
contains about $1\,056$ bytes or $8\,448$ bits, therefore, 
a theoretical maximum of $11\,837$ datagrams could be sent per second across
a 100 Mbit/s network.
Again, in practice, we find it difficult to sustain a rate of more than
$1\,500$ datagrams per second -- less than 14 percent of the maximum.

Normally, the performance of a neural network model can be measured in 
terms of the number of coupling updates per seconds (cups). For the 
conventional parallel updating algorithm we were able to
achieve on the SGI O2 workstation used to control our messaging algorithm
a speed of $5.5\cdot 10^6$ cups, while the ICMP messaging algorithm itself 
reaches $0.76\cdot 10^6$ cups.  In other words, the messaging algorithm is 
only approximately 7 times slower than a conventional algorithm running 
on a single processor. Hence, even though this algorithm is not quite as 
efficient as an
equivalent program running on a single processor, its far more efficient than
the deterministic algorithms examined here or by Barab\'asi, 
et al.\cite{barabasi01}

It should be noted, that neither the computer used for these experiments nor
the laboratory network represent state-of-the art resources.  It is possible
to obtain single processor computers an order of magnitude faster than
the one used in this study, and networks about two orders of magnitude
faster. It would be interesting then to repeat these experiments, especially
for the stochastic neural network model, on a state-of-the-art network to
determine whether or not the performance gap decreases.

Recall, that in the introduction, we talked about developing algorithms to
take
advantage of idle resources and with respect to this goal, the neural
network algorithm is successful.  Unlike the network,
the host computer can easily
sustain an output of much more than $1\,500$ datagrams per second. In order to
slow it down to this this speed, the program pauses execution after each 
packet is sent. During this pause, a timesharing
operating system can switch to another processes and perform some useful work
before returning to the neural network program.

In conclusion, the approach described herein represents an intriguing
alternative for performing parallel, distributed processing and we expect 
this approach to become more attractive with increasing network bandwidth.
Especially in the area of stochastic algorithms, the results generated to
date are promising enough to justify continued investigation.

\medskip
\section{\large\bf Acknowledgments}
\smallskip\noindent

The author would like to thank J. Klaas and F. Zimmermann for many
useful discussions related to this work.

\newpage

\bibliographystyle{unsrt}
\bibliography{bibliography}

\end{document}